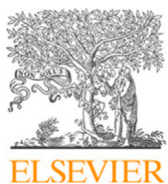
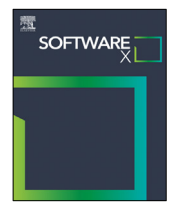

Original software publication

# Open community platform for hearing aid algorithm research: open Master Hearing Aid (openMHA)

Hendrik Kayser [a,b,c,*], Tobias Herzke [b,c], Paul Maanen [b,c], Max Zimmermann [b,c], Giso Grimm [a,b,c], Volker Hohmann [a,b,c]

[a] *Carl von Ossietzky Universität Oldenburg, Department of Medical Physics and Acoustics - Auditory Signal Processing and Hearing Devices, D-26111 Oldenburg, Germany*
[b] *Hörzentrum Oldenburg gGmbH, Marie-Curie-Str. 2, 26129 Oldenburg, Germany*
[c] *Cluster of Excellence "Hearing4all", Germany*

**ABSTRACT**

*open Master Hearing Aid* (openMHA) was developed and provided to the hearing aid research community as an open-source software platform with the aim to support sustainable and reproducible research towards improvement and new types of assistive hearing systems not limited by proprietary software. The software offers a flexible framework that allows the users to conduct hearing aid research using tools and a number of signal processing plugins provided with the software as well as the implementation of own methods. The openMHA software is independent of a specific hardware and supports Linux, macOS and Windows operating systems as well as 32-bit and 64-bit ARM-based architectures such as used in small portable integrated systems. www.openmha.org

© 2021 The Authors. Published by Elsevier B.V. This is an open access article under the CC BY-NC-ND license (http://creativecommons.org/licenses/by-nc-nd/4.0/).

## Code metadata

| | |
|---|---|
| Current code version | 4.16.1 |
| Permanent link to code/repository used for this code version | https://github.com/ElsevierSoftwareX/SOFTX-D-21-00049 |
| Legal Code License | AGPL-3.0 |
| Code versioning system used | git |
| Software code languages, tools, and services used | C++, C, Matlab/Octave, TeX, Node-RED, Python, Java |
| Compilation requirements, operating environments & dependencies | See https://github.com/HoerTech-gGmbH/openMHA/blob/master/COMPILATION.md |
| Link to developer documentation/manual | http://www.openmha.org/documentation/ |
| Support email for questions | info@openmha.org |

## Software metadata

| | |
|---|---|
| Current software version | 4.16.1 |
| Permanent link to s of this version | https://github.com/HoerTech-gGmbH/openMHA/releases/tag/v4.16.1 |
| Legal Software License | AGPL-3.0 |
| Computing platforms/Operating Systems | Linux, macOS, Microsoft Windows, Armv7, AArch64 |
| Installation requirements & dependencies | See https://github.com/HoerTech-gGmbH/openMHA/blob/master/INSTALLATION.md |
| Link to user manuals | http://www.openmha.org/documentation/ |
| Support email for questions | info@openmha.org |

## 1. Motivation and significance

The development of hearing aid signal processing methods and new algorithms for assistive listening devices is aimed at providing a benefit for the communication abilities for users in real-world scenarios. In today's hearing aids, digital audio signal processing is employed to tackle the various requirements to be met by these devices [1]. From the audiological point of view, hearing loss compensation and the related recruitment phenomenon are important items that need to be addressed to improve speech intelligibility in people with hearing impairment. A variety of algorithms, in particular automatic gain control, dynamic range compression, noise reduction as well as spatial filtering such as directional microphones of acoustic beamformers are employed to tackle this task. Further, the suppression of acoustic feedback plays an important role as well as algorithms for acoustic scene analysis and classification to adjust processing parameters or select a program in the hearing aid to provide optimal processing dependent on the communication scenario. Research in this field is carried out at universities as well as by hearing aid companies and audio equipment manufacturers. The two latter groups provide end-user devices that are developed on proprietary systems, which are not accessible to the research community and which are subject to commercial constraints. Open tools for hearing device research are a way to foster transfer of research outcomes into the design of hearing aids, cochlear implants and consumer devices for the target group's benefit. Furthermore, open tools may accelerate relevant studies with novel algorithms, lower the barriers for hardware and software development in research environments, facilitate collaborative research efforts and enable more reproducibility in hearing research. The open community platform for hearing aid algorithm research *open Master Hearing Aid* (openMHA, [2,3]) is such a tool, which has reached the level of maturity to be used in a wide research community for hearing device research and for studies with hearing impaired subjects [4–30], see Section 1.1 for an overview.

* Corresponding author at: Carl von Ossietzky Universität Oldenburg, Department of Medical Physics and Acoustics - Auditory Signal Processing and Hearing Devices, D-26111 Oldenburg, Germany.
*E-mail address:* hendrik.kayser@uni-oldenburg.de (Hendrik Kayser).

An important part of the development of new methods is their evaluation. The first step in the evaluation of novel methods is usually simulations that are carried out in an offline manner, i.e., all audio data is processed in a batch and causality is not necessarily met. This first step typically tests the performance of a method in isolation from other operations that are part of the signal processing chain in a real device. The outcome of such a simulation is an important result in terms of the capabilities and the potential performance of the method itself. However, it is not a good predictor of the benefit the method can provide for a user in real-life conditions. For this, some important further aspects need to be taken into account: The system needs to be considered as a whole, i.e., the signal processing algorithm needs to work in combination with all other components of the system to provide an estimate of benefit in a realistic usage scenario. This "holistic" approach should include essential hearing aid functionality such as amplification and dynamic range compression to compensate for hearing loss. Feedback suppression techniques to minimize howling caused by acoustic feedback loops between hearing aid loudspeaker and microphone also play a role to allow for sufficiently high amplification of the incoming audio signal. The coupling of a hearing device to the ear affects the lower limit of sound level that can be present at the eardrum and also determines the ratio between unprocessed sound signals arriving directly and sound processed by the hearing device. In addition, signal enhancement methods are used to increase speech intelligibility, e.g., directional microphones, single- and multi-channel noise reduction algorithms. All these processing steps underlie two important technical constraints: They must be accomplished in (a) real-time[1] and (b) with a low latency[2] between input signal at the hearing aid microphone and playback at the hearing aid loudspeaker. This is required to avoid an echo effect and to provide natural perception of the user's own voice and synchronicity between visual cues such as lip movements and incoming sound signal, thus enabling natural communication in real-world scenarios. A latency of 10 ms is a value that is typically defined as the upper limit for the latency [32].

All the aspects listed so far should be taken into account in advanced simulations of the hearing aid processing chain. This means, the processing chain needs to be run live on suitable hardware.

Furthermore, to enable differently-skilled target groups to use the software tools for research, several approaches to realize and modify hearing aid processing are required. These include the implementation of new signal processing algorithms using an established programming language, the possibility to set up and configure hearing aid processing chains based on the available components without knowledge of a programming language as well as the possibility to control available setups using high-level user interfaces and tools without detailed knowledge of the software itself.

openMHA is a comprehensive software package that meets these requirements and provides a modular framework to set up a hearing system with all required processing steps. It originates from the commercial closed-source software *Master Hearing Aid* [31] that was made available and is further developed under open-source license. It features mechanisms to ensure real-time-safe audio signal processing,[3] real-time-safe configuration changes at runtime,[4] and monitoring of variables during processing such that a hearing aid algorithm chain under development and test can be easily monitored and validated. In addition, beyond providing the fundamental structure for hearing aid processing, the software package includes a large and growing set of signal processing plugins that can be used to create a multitude of configurations without requiring knowledge of the C/C++ programming language. New developments can be integrated using openMHA's algorithm development framework using C/C++ or the Matlab Coder [33].

A scientific problem that is tackled with openMHA is the realization of more ecologically valid studies in the context of hearing aid research with the aim to increase the degree to which laboratory-based research findings reflect real-life hearing-related function. The crucial aspect is that researchers are enabled to investigate, implement and test their developments in a realistic hearing aid processing chain using realistic acoustic signal scenarios.

### 1.1. Usage of openMHA in the research community

Since its first release in 2017 openMHA is utilized in research related to hearing aids, evaluation of signal processing methods and for the setup of hearing aid processing in real-time systems.

---

[1] In the context of audio signal processing in openMHA, *real-time* processing means that a fragment of the incoming audio signal can be processed within the temporal duration of the respective fragment.

[2] *Latency* refers to the time which passes between the input of an incoming audio sample and the output of that sample after processing. Latency is influenced by the properties of the audio hardware, the size of the signal fragments used for processing as well as signal buffering as required, e.g., in STFT frameworks, see [31] for more details. Note that algorithmic delay such as, e.g., the group delay of filters in the signal processing chain, add to the latency defined here.

[3] Audio signal processing can be performed in dedicated threads with high priority settings. Furthermore, the audio signal processing implemented in openMHA avoids that can introduce unpredictable delays (e.g., memory allocation, file system access).

[4] Changes of configuration parameters of signal processing algorithms while signal processing is continuously ongoing will not induce additional adaptation operations within the signal processing thread, and therefore do not increase the computation time required for signal processing. This is achieved by performing all required preparations for parameter changes outside the real-time signal processing thread, typically with lower priority settings.

The software has been used for various purposes such as fitting of a multi-band dynamic range compressor to the hearing loss of hearing-impaired subjects a) in order to model speech intelligibility in cochlear implant users [4], predicting speech recognition performance in aided subjects [5] also including a directional microphone available with openMHA, and evaluation of near-end listening enhancement [6] with individual hearing loss compensation. In that context the openMHA fitting tool (see Fig. 2) and the gain prescription rules implemented therein were employed to realize the hearing aid fitting procedure. Perceived sound quality for music-listening and speech with hearing aids and preferred gain settings were evaluated [7–9]. A study on the impact of movement behavior of human subjects on the potential benefit provided by different hearing aid processing methods [10,11] used a set of reference implementations of algorithms available with openMHA for a post-analysis that incorporates movement trajectories of the head and eyes in the dynamic virtual acoustic scenes. This work lead to an open-access database [12] that contains the measured trajectories and parameters of the scenes. A software for rendering of acoustic scenes [13] that supports the integration and control of openMHA from within the software is used together with openMHA to enable the reproduction of these virtual scenes by the user.

openMHA was used to realize real-time signal processing for an ear-level device for hearing aid research [14], which was compared to commercially available devices in terms of a technical [15] and a perceptual [16] evaluations.

Several studies used signal processing algorithms implemented in openMHA as references in the evaluation of new developments or to complement other processing methods. A recurrent neural network (RNN) was used to perform noise reduction and dereverberation and compared to the available single-channel noise reduction (SCNR, [34]) algorithm [17]. Furthermore the author implemented the RNN-based processing as an openMHA plugin in order to use it on portable hardware. In a different study, the same SCNR algorithm was found to be superior compared to a statistical-model-based noise reduction algorithm for stationary background noise and at poor SNRs [18]. Furthermore, the combination of our SCNR method with additional envelope enhancement turned out to be beneficial. In subject experiments with normal hearing and unilateral implanted CI listeners that evaluated directional microphone algorithms for speech enhancement in these target groups [19], openMHA was used to conduct the hearing loss compensation and the spatial filtering. The multi-band dynamic compressor and filterbanks provided with our software were integrated in a bone conduction hearing aid [20]. The *Clarity Challenge* project [21], which organizes machine learning challenges for hearing devices on speech enhancement with hearing aid signal processing and perception models of speech intelligibility, uses an openMHA configuration as a baseline hearing aid. It features adaptive differential microphones [35] and provides hearing loss compensation based on the Camfit compressive gain prescription rule [36]. openMHA was integrated with the Julia programming language [22] to implement Bayesian machine learning techniques for real-time hearing aid processing. The real-time capabilities of openMHA were also used in an computer vision application for real-time feature extraction and classification of posture and gesture [23].

openMHA is also suitable to conduct hearing aid signal processing on portable integrated systems that are enabled by the availability of small hardware with sufficient computational power such as single-board computers (SBC) or systems on a chip (SoC). Such hardware platforms are of particular interest to conduct hearing aid research outside the laboratory. For that purpose, openMHA was ported to an open-source SoC Field Programmable Gate Array (FPGA) platform [24–26]. Integrated portable platforms that feature ear-level devices such as behind-the-ear devices are of particular interest. In combination with a software suite that provides signal processing capabilities as well as measurement applications for hearing research, such as openMHA, a complete field test system is available. See Section 3 for an example of such a field test setup that is tailored to hearing device research with openMHA [27]. It was found to have "performed on a comparable level as the commercial hearing aids for the assessed metrics and thus can be viewed as a fully functional hearing aid for research purposes" [28]. In a recent study [29], this portable platform was extended with a mobile EEG acquisition system aiming at utilizing brain signals for hearing aid processing. Another portable hearing aid research platform [30] has been shown to be capable of running openMHA, which was included in the software suite for its processing board [37]. In summary, the large amount of studies performed at different research institutions by researchers with different background using the openMHA for different research questions shows that the openMHA software package in its current version is fully appropriate.

### 1.2. User groups

To optimally support the community, we divided the user group into three different levels that we address in the documentation and tools that are provided with the software package.

*Audiological researcher.* This user group carries out audiological research studies with hearing-impaired and normal-hearing subjects. For this purpose configurations and graphical user interfaces (GUI) are provided as tools with the openMHA software. Examples for these tools are a GUI for hearing aid fitting that runs under Matlab and Octave and a webserver-based GUI.

*Application engineer.* The group of application engineers builds setups for experiments and studies to be performed by audiological researchers. This group uses the features and tools that are provided with the openMHA software package. It contains a number of signal processing plugins that can be combined to complete hearing aid signal processing chains using a text-based configuration language. GUIs or other control applications can be created, for instance, in Matlab or Octave, in Python, or in Node-RED [38].

*Plugin developer.* In case that a functionality needed for a particular setup is not available in its current version such as a newly developed signal processing algorithm, the software can be extended with own plugins. A developer uses C/C++ programming language to implement new features into the framework. A large library of signal processing methods to draw from is included with the openMHA software development framework. Furthermore, the Matlab Coder [33] can be used to implement new openMHA plugins or to compile a signal processing library that can be loaded by a specific wrapper plugin in openMHA. In the latter case, new plugins can be developed and implemented without any knowledge of C++ and without having an openMHA development environment available for compilation.

Manuals and examples of documented source code are provided, see end of Section 2.3.3 for more details.

### 1.3. Available algorithms

The software package includes plugins for basic operations such as calibration, filtering, resampling, amplification and an overlap-add Fourier analysis and synthesis framework for signal processing in the spectral domain. Furthermore a number of plugins are included that cover processing methods of the following types: multi-band dynamic range compression [39], adaptive feedback cancellation [40], directional microphones [35], binaural noise and feedback reduction [41], binaural beamforming [42–44], single-channel noise reduction [34], and sound source localization [45]. In some cases ([34,35,41,42], adaptive beamforming [44], and a delay-and-subtract beamformer), reference configurations are provided that have been used the same way in a research study with the aim to enable reproducibility of the experimental setup by other researchers.

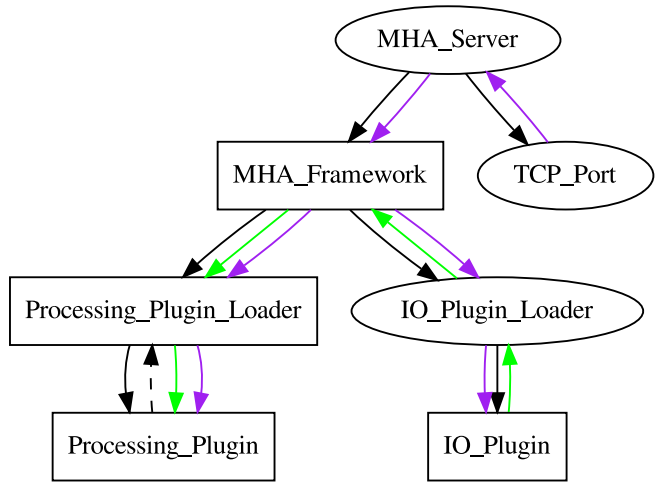

**Fig. 1.** openMHA software architecture visualization. Black arrows: object ownership. Dashed black arrow: Some plugins contain themselves another Processing_Plugin_Loader instance by which they can load child plugins. Green arrows: audio signal path. Audio signal can have arbitrary number of audio channels. Arrow direction depicts incoming signal, the processed output signal travels in the reverse direction along the same path. Purple arrows: Configuration command execution path. Responses to configuration commands travel along the same path in reverse direction. Rectangular objects employ the configuration language parser to interpret or propagate configuration commands. (For interpretation of the references to color in this figure legend, the reader is referred to the web version of this article.)

## 2. Software description

This section presents functionalities of openMHA (2.1), software architecture (2.2), code examples (2.3) for different user scenarios, and our practices and tools used for the software development (2.4).

### 2.1. Software functionalities

openMHA processes audio signals with hearing aid signal processing algorithms implemented as plugins. Using the configuration language, different plugins can be combined and configured to form the desired complete signal processing chain.

openMHA can process either live audio signal from a sound card, or it can process audio input signal from input sound files and produce output sound files. When processing a live audio signal, the configuration of individual plugins can be altered without interrupting the audio processing, making it possible to subjectively compare different algorithm configurations.

Different signal processing plugins are available that import and export data to and from files in different formats and over the network (open sound control (OSC, [46]), lab streaming layer (LSL, [47])). The latter is useful to take information from other sensors – such as head- and eye motion and EEG data – into account in audio signal processing methods. To this end, plugins can be implemented that analyze such data in order to control signal processing in a certain way. Examples are the exploitation of head- and eye movement to steer directional filtering, EEG data for auditory attention decoding or listening effort.

### 2.2. Software architecture

The overall software architecture is depicted in Fig. 1. In the following, the single components are described in detail.

#### 2.2.1. openMHA server

openMHA is mainly implemented in C++ as a command line application mha (MHA_Server in Fig. 1) without a graphical user interface. The command line application processes its command line parameters, then opens a TCP port (TCP_Port) and listens for incoming configuration commands. The MHA_Server mha creates one instance of the openMHA framework (MHA_Framework) class which functions as the main event dispatcher of openMHA. The openMHA framework reacts to configuration command events by either

(1) interpreting framework configuration commands itself with the help of the MHA configuration language parser, or
(2) forwarding configuration commands affecting the audio IO plugin to the respective IO plugin, if one is loaded, or
(3) forwarding configuration commands affecting one of the signal processing commands to the processing plugin loader.

The openMHA framework can load a single IO plugin (IO_Plugin) and a single signal processing plugin (Processing_Plugin) with the help of the respective plugin loaders, the IO plugin loader (IO_Plugin_Loader) or the signal processing plugin loader (Processing_Plugin_Loader).

#### 2.2.2. Plugin_Loaders and plugins

The plugin loaders search for MHA plugins in a configurable set of file system directories and load the first matching openMHA plugin into the process. Newly loaded plugins insert their configurable settings into a tree of configuration settings, and they are also inserted into the signal processing path.

IO plugins as well as signal processing plugins are compiled into shared libraries which export a short list of functions, i.e., their interface. The respective interface's shared libraries are loaded into the running MHA_Server application mha, which in turn are used by the plugin loaders to integrate the plugin into the mha.

Classes and functions shared between the MHA_Server application mha and the plugins are compiled into a shared library, the openMHA Toolbox Library libopenmha.

*IO plugins.* IO plugins connect the openMHA to audio data sources/ sinks like sound cards or sound files. openMHA does not impose any restrictions on audio sampling rates, number of audio channels, etc.

*Signal processing plugins.* Signal processing plugins implement hearing aid signal processing algorithms. Some of the existing signal processing plugins contain processing plugin loader(s) themselves and can load other signal processing plugin(s), creating a tree of processing plugins.

#### 2.2.3. Configuration language interpreter

The MHA configuration language interpreter is used by the openMHA framework, the IO plugins, the signal processing plugin loader and the signal processing plugins. It is contained in the openMHA Toolbox library, a shared library linked by the MHA server as well as by the individual plugins. It provides a text-based interface to a tree of configuration variables of different data types. Configuration variables are declared by the framework, the plugins, etc. during their respective initialization, registered with the configuration language tree and made accessible by the openMHA server mha to command line arguments and to commands received on the TCP port. The configuration language interpreter can invoke registered C++ callbacks when a configuration variable is read from or written to, thereby enabling configuration changes before and during signal processing.

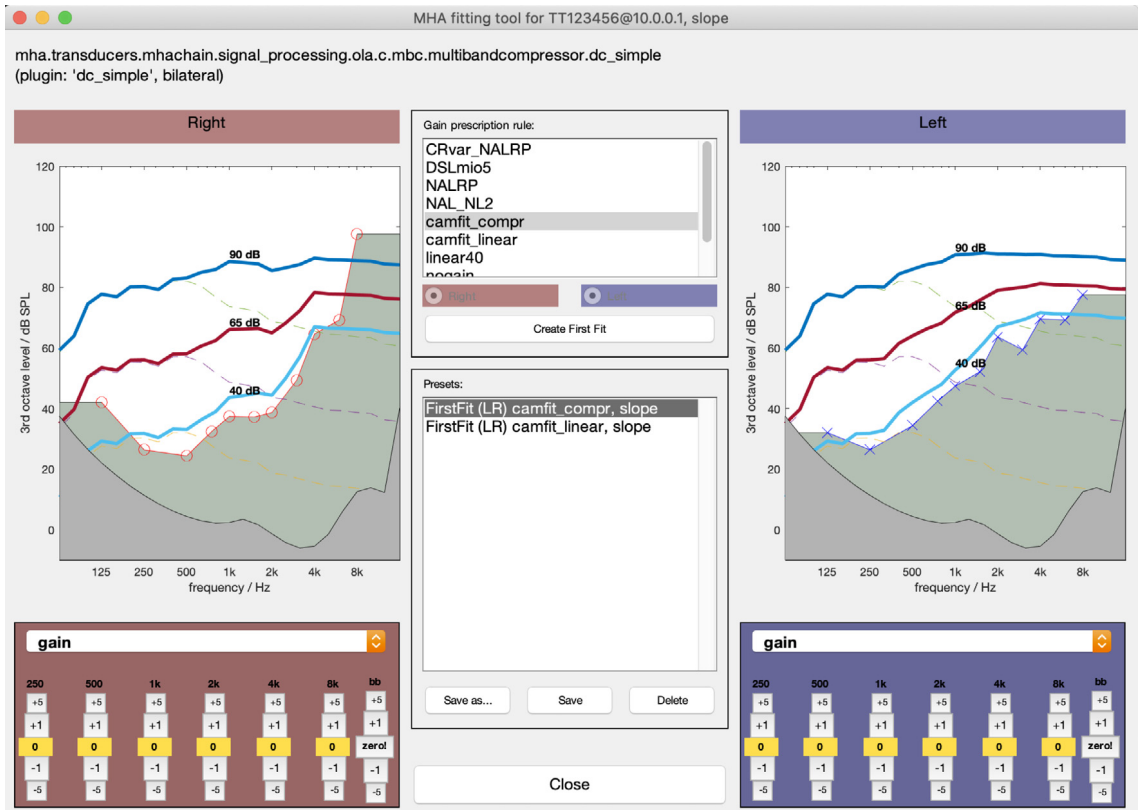

**Fig. 2.** Graphical user interface of the openMHA Octave/Matlab fitting tool.

The design decision to develop a custom configuration language was taken to allow for a lightweight language interpreter, as required for minimal portable implementations, which still allows for the flexibility of extensions through pre- and post-assignment hooks for seamless interaction between a configuration process and the dropout-free real-time signal processing.[5] The MHA configuration language is similar to Matlab/Octave language. A detailed description can be found in [31] and in the openMHA Application Manual, an example is provided in Section 2.3.2.

### 2.2.4. Configuration tools

Independent software tools can be used to interact with the openMHA using the TCP interface. We provide Matlab/Octave tools for interacting with a running openMHA server over the TCP connection ranging from functions for retrieving and setting MHA configuration variables to graphical user interfaces (GUI) for different tasks, e.g. calibration or adapting a dynamic compressor to individual hearing losses. These tools can execute on the same computer as the openMHA server mha or can also execute on a different computer and communicate with the openMHA server over a wired or wireless network. For embedded systems and smartphones, Matlab or Octave are not a feasible solution to control an openMHA instance on the same or a remote device. In such a case, an embedded webserver that runs on the processing platform itself and provides a GUI for any connected browser-enabled device is a universal approach. An example that uses Node-RED [38] is shown in Fig. 4 in Section 3.

### 2.3. Sample code snippets analysis

In the following, one exemplary code sample is shown for each user group described in Section 1.2. Section 2.3.1 depicts how to start an available openMHA configuration file, which does not require any interaction with the configuration language or the source code itself. In Section 2.3.2 the structure of the configuration used in Section 2.3.1 is described in detail. Understanding such a configuration file enables the user to modify plugins and parameters in available configurations and create own configurations using the plugins that are available with openMHA. Section 2.3.3 describes the development of a simple signal processing plugin on source-code level. These examples and many more can be found in the openMHA software package. In combination with the manuals and tutorials available with the openMHA documentation, the examples presented here can be used as a basis for the users' own developments.

### 2.3.1. Invocation and control

Audiologists can invoke openMHA with existing configurations to perform experiments. The example configuration from the following subsection – a realization of multi-band dynamic range compressor – can be invoked from the command line, e.g., as

```
mha ?read:examples/14-dc-
simple/example_dc_simple_live.cfg cmd=start
```

---

[5] High-level languages such as Python, lua or ruby were considered for the configuration language. However, since primarily only low-level configuration functionality is required in openMHA, they would exceed the requirements of openMHA by far.

To control an experiment from Matlab or Octave, openMHA can also be started with the same configuration from Octave:

```
mha = mha_start(); % start openMHA process
mha_query(mha,'read','examples/14-dc-
simple/example_dc_simple_live.cfg');
mha_set(mha,'cmd','start');
mhacontrol(mha); % start Control and Fitting GUI
```

The last command will open a graphical user interface with access to level meters, calibration dialog, profiling tools as well as an audiogram database and a fitting user interface as shown in Fig. 2 to apply different gain prescription rules and to modify gain parameters in the running instance of openMHA. A detailed description of this tool is available in the Fitting GUI manual. It provides a number of gain prescription rules and also offers the possibility to implement own gain rules in Octave/Matlab based on a template available with the software.

openMHA can also be used to batch-preprocess sound files with hearing aid dynamic compression for specific hearing losses:

```
offline_fitting_tool(); % starts a wizard that guides users
```

The command above opens a GUI-based Matlab/Octave tool for processing a large number of audio files with hearing aid signal enhancement adapted to custom hearing losses. The tool guides the user through the selection of sound files, audiogram, and a fitting rule with the same interface.

*2.3.2. Configuration file example*

Application engineers can use a simple text-based configuration language to describe which plugins to load, how to arrange and configure them. The following example – the configuration invoked in Section 2.3.1 – performs multi-band dynamic compression of a binaural input signal, where the compression is performed in the Short-Time Fourier Transform (STFT) domain:

```
1  # Number of input audio channels
2  nchannels_in = 2
3  # Signal fragment size in samples
4  fragsize = 64
5  # Sampling rate
6  srate = 44100
7  # MHA library name
8  mhalib = transducers
9  # IO plugin library name
10 iolib = MHAIOJack
11 # Select audio input and output ports
12 io.con_in = [system:capture_1 system:capture_2]
13 io.con_out = [system:playback_1 system:playback_2]
14 # Load STFT framework plugin
15 mha.plugin_name = overlapadd
16 # Calibration settings
17 mha.calib_in.peaklevel = [116 116]
18 mha.calib_out.peaklevel = [114 114]
19 # STFT parameters
20 mha.overlapadd.fftlen = 256
21 mha.overlapadd.wnd.len = 128
22 # Load mhachain for spectraldomain processing
23 mha.overlapadd.plugin_name = mhachain
24 # load plugins. dc_simple performs dynamic compression.
25 mha.overlapadd.mhachain.algos = [fftfilter-
bank dc_simple combinechannels]
26 # Frequency bands
27 mha.overlapadd.mhachain.fftfilterbank.f = [250 1000 4000]
28 # Threshold of noise gate in dB SPL
29 mha.overlapadd.mhachain.dc_simple.expansion_threshold
   = [20 20 20 20 20 20]
30 # Slope of level mapping below noise gate
31 mha.overlapadd.mhachain.dc_simple.expansion_slope
   = [4 4 4 4 4 4]
32 # Gain at 50 dB SPL
33 mha.overlapadd.mhachain.dc_simple.g50 = [10 25 40 11 31 55]
34 # Gain at 80 dB SPL
35 mha.overlapadd.mhachain.dc_simple.g80 = [5 15 10 5 21 19]
36 # Limiter threshold, a.k.a maximum possible out-
put level, in dB SPL
37 mha.overlapadd.mhachain.dc_simple.limiter_threshold
   = [120 120 120 120 120]
38 # attack time constant in s
39 mha.overlapadd.mhachain.dc_simple.tau_attack = [0.02]
40 # decay time constant in s
41 mha.overlapadd.mhachain.dc_simple.tau_decay = [0.1]
42 # Name of fftfilterbank plugin. Used to extract frequency in-
formation.
43 mha.overlapadd.mhachain.dc_simple.filterbank = fftfilterbank
44 # Number of audio channels for re-combination after filterbank
45 mha.overlapadd.mhachain.combinechannels.outchannels = 2
```

This configuration is available in the openMHA example 14, "dc_simple". It implements a simple multi-band dynamic range compression scheme. In lines 1–13 the basic parameters such as number of audio channels, blocksize in samples, sampling rate, and hardware-dependent connections are set. The Jack Audio Connection Kit (JACK) [48] is employed for audio in- and output and the according library is loaded. As the first plugin (mhalib) the transducers plugin is loaded that handles calibration. Lines 14–21 set up the STFT framework (overlapadd) parameters and values that affect the calibration in openMHA which is crucial for dynamic compression. In lines 22–45 the actual processing chain for the compressor is configured including a filterbank that splits the signal into the target frequency bands, the gain values for the compressor and time constants. See comments in the code given above for more details.

General information on the configuration language can be found in the openMHA Application Manual.

*2.3.3. Plugin development example*

Algorithm developers can develop a new plugin by inheriting from the openMHA class MHAPlugin::plugin_t<> and implementing a few methods. The following code implements a plugin that attenuates sound by 20 dB:

```
#include "mha_plugin.hh"
class attenuate20_t : public MHAPlugin::plugin_t<int> {
public:
  attenuate20_t(algo_comm_t & ac,
          const std::string & chain_name,
          const std::string & algo_name)
    : MHAPlugin::plugin_t<int>("This plugin attenuates by 20dB"
,ac)
  {}
  void release(void) override
  {}
  void prepare(mhaconfig_t & signal_info) override
  {
   if (signal_info.domain != MHA_WAVEFORM)
    throw MHA_Error(__FILE__, __LINE__,"
can only process waveform");
  }
  mha_wave_t * process(mha_wave_t * signal)
  {
   // -20dB = factor 0.1
   MHASignal::for_each(signal,[](mha_real_t sample){return sam-
ple * 0.1f;});
   return signal;
  }
};
MHAPLUGIN_CALLBACKS(attenuate20,attenuate20_t,wave,wave)
MHAPLUGIN_DOCUMENTATION(attenuate20, "
example level-modification",
 "Plugin attenuate20 attenuates the input signal by 20dB.")
```

The new plugin class needs to implement the constructor and the methods prepare(), process(), and release(). The Plugin development guide contains a plugin development tutorial as well as a reference of classes and methods of the openMHA toolbox library that can be used while developing signal processing algorithms with the openMHA. This example plugin is available in openMHA as "attenuate20".

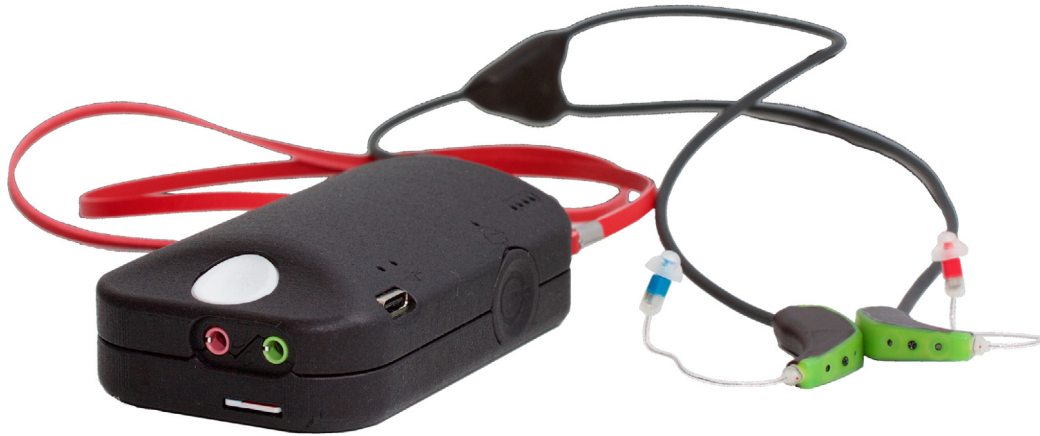

**Fig. 3.** The "Portable Hearing Laboratory" (PHL, [27]).

Alternatively, the Matlab Coder [33] software is supported in openMHA. This functionality enables the conversion of Matlab code into openMHA-compatible C/C++ code or libraries that can be loaded into a processing chain using a wrapper plugin such that knowledge of C/C++ is not required if a user wants to implement an own Matlab-based signal processing method in openMHA. Researchers can follow and adapt a complete example for producing an openMHA plugin from Matlab code, general documentation is available in the Matlab Coder Integration manual. When following this procedure, it is also possible to split the process by performing the Matlab-to-C++ conversion on a different computer (with Matlab Coder available) than the compilation of the generated C++ code into an MHA plugin (with the compilers for the desired target platform available).

*2.4. Development procedure and practices*

Users can submit issues, feature requests and pull requests through the openMHA GitHub repository [49]. Users can post questions about openMHA usage in the openMHA forum [50] where they are usually answered within a few days by other users or by the openMHA development team.

The openMHA development team uses a non-public Phabricator [51] instance for internal issue tracking, feature requests, peer code reviews, democratic decisions, and documentation of development processes.

We use git [52] for version control. Features and bug fixes are developed in dedicated branches and merged into the public development branch only after peer code reviews. We are using unit tests written with Googletest/Googlemock [53] to test details of our openMHA application, our toolbox library, or individual openMHA plugins. Additionally, we use system tests written in Matlab/Octave to set up complete instances of openMHA processing configurations and to test the effects of signal processing by individual openMHA plugins or by combinations of several openMHA plugins against expectations. These tests are integrated into our build process which uses GNU Make [54] and can be invoked with "`make unit-tests`" for the unit tests or "`make test`" for the system tests, respectively. Additionally, we use a Jenkins [55] server to automate recompilation and execution of the aforementioned unit tests and system tests on all openMHA target platforms (Windows, macOS, Linux for PC, Linux for ARM) after every update of the development branch, and also for every suggested code review request. Code review requests will only be sent to reviewers once all builds and tests have succeeded on all openMHA target platforms. We use the GNU GCC compilers on Linux [56] and Windows [57] and the Clang compiler on macOS [58] to compile openMHA, and have configured these compilers to treat all warnings as errors (`-Wall -Werror`) in order to force ourselves to fix all code smells detected by modern C++ compilers. The minimum C++ standard that openMHA supports is C++17.

To release new openMHA versions, the openMHA development branch on git is merged into the master branch on git, but only after additional live audio processing tests have been performed by the openMHA development team and which cannot be automated.

**3. Illustrative example**

One example of how openMHA can be used in the field is the "Portable Hearing Laboratory" (PHL, [27]), see Fig. 3. The PHL consist of a *BeagleBone Black wireless* single-board computer that was extended with a multi-channel audio board based on the open-source audio interface Cape4all [59]. In addition to the processing board and audio interface the integrated system includes a battery and a set of binaural behind-the-ear (BTE) hearing aids that comprise two microphones on each side to capture the sound field and receivers in the canal (RIC) for the playback of processed audio signals. The light-weight system can be worn by a user using a neckband, the BTE-RIC devices are connected with the processing box via flexible cables. An optimized Linux operating system - *MAHALIA* [60] - is available that is tailored to performing low-latency signal processing on the processing board.

A typical use case for the PHL is a field test of a new hearing aid fitting strategy or gain prescription rule. As it is not possible to implement a new fitting rule on the hearing aid of a hard-of-hearing subject their device can be replaced by the PHL for the experiment. Input-level-dependent gain values can be pre-computed according to the prescription rule under test taking into account audiometric measures such as hearing threshold and loudness perception or directly modified at runtime on the device using the fitting tool shown in Fig. 2. The resulting values are effective in the multi-band dynamic range compressor plugin which is available in openMHA. This way, the subject has an individualized hearing aid that can be worn and tested in everyday situations. For the evaluation, e.g., questionnaires can be used or feedback can be obtained in interviews with the participant. Furthermore, the subject may be given access to certain parameters of the processing to adjust the device in different situations or compare different processing modes. Therefore, a smartphone (or any other browser-enabled device with wireless network functionality) can be connected to the PHL device through a wireless network connection. A webserver running on the PHL can be used to provide a GUI as shown in Fig. 4.

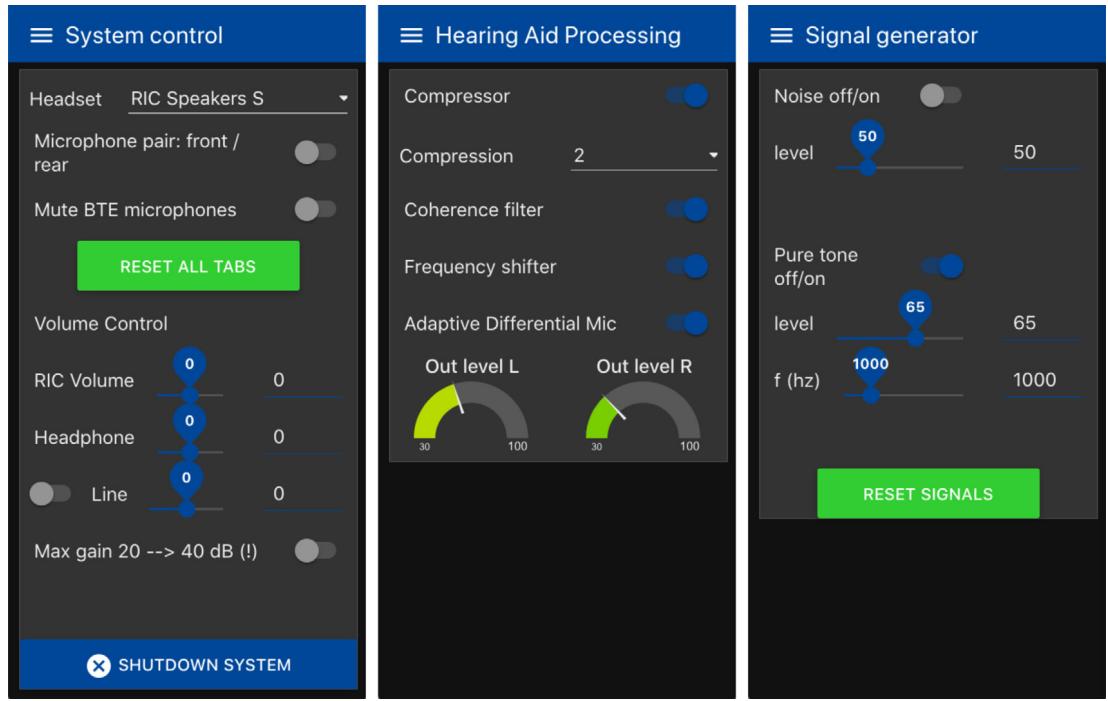

**Fig. 4.** GUI running on the Portable Hearing Laboratory (PHL), containing three different tabs. The interface is accessible through a web browser from a smartphone, tablet or computer connected to the PHL via wifi. The left panel *System control* provides access to headset, microphones and gain settings. The center panel *Hearing Aid Processing* is used to control the hearing aid processing — activation/deactivation of single processing stages and modification of the compression ratio. The resulting broadband output level is monitored at the bottom. The right panel *Signal generator* allows to generate noise and pure tone signals in the hearing aid with controllable level and frequency.

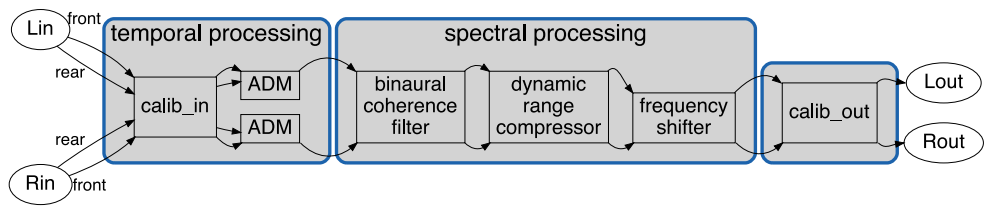

**Fig. 5.** Hearing aid signal processing chain provided with the Portable Hearing Laboratory (PHL) system software MAHALIA.

This particular GUI provides control over a basic hearing aid configuration that is provided with the MAHALIA operating system sd-card image (available from https://mahalia.openmha.org/) and also available in the openMHA configuration examples as example 17 "PHL-generic-hearing-aid". The configuration provides a hearing aid processing chain as sketched in Fig. 5. It uses the two microphone signals from each of the BTE devices (*Lin* and *Rin*) as input, followed by the microphone calibration stage. For noise reduction it features directional filtering with bilateral adaptive differential microphones (ADM). The resulting binaural signal is further processed in the spectral domain with a binaural coherence filter for noise suppression and feedback reduction, a 5-band dynamic range compressor with selectable compression ratio, and a band-limited frequency shifter for feedback reduction. Finally, the signals are transformed back into the time domain, calibrated to account for receiver and ear-canal frequency response, and played back through the left and right hearing aid RIC (*Lout* and *Rout*). In addition to the BTE-RIC devices, the 3.5 mm jack connectors of the PHL can be used for audio input (left (pink) jack connector in Fig. 3) and audio output (rigth (green) jack connector in Fig. 3). The latency between input and output of the processed audio amounts to 10 ms, the audio sampling rate is 24 kHz.

The signal processing with this configuration uses 50% of the CPU capacity of the system with all processing stages activated. This leaves room for modifications of the setup in several ways, e.g., a larger number of frequency bands in the compressor, optimization towards lower latency, or extension with other signal enhancement approaches.

## 4. Impact

openMHA provides a modular framework that enables researchers to set up a flexible defined hearing system that provides full control and access to the processed signal at each processing step. This allows researchers to assess the benefit of hearing aid processing methods in several ways, ranging from offline simulations based on a set of audio files, over simulated realistic virtual audio-visual environments, to field studies in uncontrolled real-world environments. During the whole process the same software and configurations can be used with only minimal adaptation required such as switching from processing of audio files to processing of microphone input. In particular, due to the real-time capabilities of openMHA, hearing aid processing with live sound input and output at low latency can be achieved. In general, this does not require the re-implementation of the processing on a different hardware/software system, i.e., rapid prototyping with tests in realistic test scenarios become possible with openMHA. This versatility facilitates the development and evaluation of new methods with a focus of ecological validity, i.e., to increase the level to which research findings reflect real-life hearing-related function [61]. Furthermore, openMHA facilitates the investigation of interactions among a hearing aid processing strategy and the user needs.

We have identified several publications (see Section 1.1) originating from studies that actively use openMHA in the context of

- hearing aid fitting
- modeling and prediction of hearing aid benefit
- signal processing for assistive hearing devices
- speech enhancement methods including deep machine learning
- psychoacoustic experiments on hearing aid processing
- assessment of user behavior in subject-in-the loop experiments

Reproducible and accessible science is another important factor that is supported with the openMHA software. openMHA software setups that have been used in hearing studies can be made available for direct reproduction by other researchers. Some of them have already been included in the openMHA distribution, e.g., the openMHA configurations used by [11,44]. Setups can be passed between different research sites or multi-center studies can be carried out using a common code basis such as is the case in the *Clarity Challenge* [21] on machine learning for hearing devices and prediction of hearing aid benefit. In summary, the already large impact of the openMHA software on hearing-aid research is clearly demonstrated by large amount of published studies that used openMHA.

## 5. Conclusions

In this publication we introduced *open Master Hearing Aid* (openMHA), an open-source software platform for hearing aid research. The software enables sustainable and reproducible research for hearing instruments and supports studies that aim for a high level of ecological validity, i.e., that increase the degree to which research findings reflect real-life hearing-related function. It offers several services for hearing aid researchers such as fundamental functionality for hearing aid signal processing like acoustic calibration and filterbanks, a library for common signal processing tasks and a complete set of hearing aid signal processing reference algorithms, capabilities of offline-processing as well as real-time signal processing with a reliable low acoustic latency. A versatile tool set enables researchers to conduct audiological studies to investigate hearing aid processing methods 1) based on available reference configurations and control tools 2) based on custom hearing aid signal processing chains that can be configured based on a large set of available algorithms and 3) using novel algorithms that can be implemented by a user using C/C++ programming language or based on Matlab code. The software can be used on a wide range of hardware including small portable devices and under all established operating systems.

It has demonstrated its usability in a number of studies from different laboratories that used openMHA.

**Declaration of competing interest**

The authors declare that they have no known competing financial interests or personal relationships that could have appeared to influence the work reported in this paper.

**Acknowledgments**

Research reported in this publication was supported by the National Institute On Deafness And Other Communication Disorders of the National Institutes of Health, USA under Award Number R01DC015429. The content is solely the responsibility of the authors and does not necessarily represent the official views of the National Institutes of Health.